# Contribution of the Open Access modality to the impact of hybrid journals controlling by field and time effects

Pablo Dorta-González [1,*] and María Isabel Dorta-González [2]

[1] Universidad de Las Palmas de Gran Canaria, TiDES Research Institute, Campus de Tafira, 35017 Las Palmas de Gran Canaria, Spain. E-mail: pablo.dorta@ulpgc.es

[2] Universidad de La Laguna, Departamento de Ingeniería Informática y de Sistemas, Avenida Astrofísico Francisco Sánchez s/n, 38271 La Laguna, Spain. E-mail: isadorta@ull.es

* Corresponding author

**Abstract**

**Purpose:** Researchers are more likely to read and cite papers to which they have access than those that they cannot obtain. Thus, the objective of this work is to analyze the contribution of the Open Access (OA) modality to the impact of hybrid journals.

**Design / methodology /approach:** The "research articles" in the year 2017 from 200 hybrid journals in four subject areas, and the citations received by such articles in the period 2017-2020 in the Scopus database, were analyzed. The hybrid OA papers were compared with the paywalled ones. The journals were randomly selected from those with share of OA papers higher than some minimal value. More than 60 thousand research articles were analyzed in the sample, of which 24% under the OA modality.

**Findings:** We obtain at journal level that cites per article in both hybrid modalities (OA and paywalled) strongly correlate. However, there is no correlation between the OA prevalence and cites per article. There is OA citation advantage in 80% of hybrid journals. Moreover, the OA citation advantage is consistent across fields and held in time. We obtain an OA citation advantage of 50% in average, and higher than 37% in half of the hybrid journals. Finally, the OA citation advantage is higher in Humanities than in Science and Social Science.

**Research limitations:** Some of the citation advantage is likely due to more access allows more people to read and hence cite articles they otherwise would not. However, causation is difficult to establish and there are many possible bias. Several factors can affect the observed differences in citation rates. Funder mandates can be one of them. Funders are likely to have OA requirement, and well-funded studies are more likely to receive more citations than poorly funded studies. Another discussed factor is the selection bias postulate, which suggests that authors choose only their most impactful studies to be open access.

**Practical implications:** For hybrid journals, the open access modality is positive, in the sense that it provides a greater number of potential readers. This in turn translates into a greater number of citations and an improvement in the position of the journal in the rankings by impact factor. For researchers it is also positive because it increases the potential number of readers and citations received.

**Originality /value:** Our study refines previous results by comparing documents more similar to each other. Although it does not examine the cause of the observed citation advantage, we find that it exists in a very large sample.

## Introduction

Researchers are more likely to read and cite papers to which they have access than those that they cannot obtain. Thus, since the emergence of the world wide web, scientists and scholarly publishers have used different forms of Open Access (OA), a disruptive model for the dissemination of research publications (Björk, 2004). In the last years, more and more scientists are making their research results openly accessible to increase its visibility, usage, and citation impact (Dorta-González et al., 2017; 2020).

The common characteristic of all different forms of OA is that the primary source of communication of research results, the peer reviewed article, is available to anybody with Internet access free of charge and access barriers (Prosser, 2003).

There are four main OA modalities. *Gold OA* refers to scholarly articles in fully accessible OA journals. *Green OA* refers to publishing in a subscription or pay-per-view journal, in addition to self-archiving the pre- or post-print paper in a repository (Harnad et al., 2004). *Hybrid OA* is an intermediate form of OA, where authors pay scholarly publishers to make articles freely accessible within journals, in which reading the content otherwise requires a subscription or pay-per-view (Björk, 2017). And finally, *Delayed OA* refers to scholarly articles in subscription journals made available openly on the web directly through the publisher at the expiry of a set embargo period (Laakso & Björk, 2013).

As previously said, a hybrid journal is a traditional one, for which readers need a subscription or where readers can pay to view individual articles. However, the journal offers authors the possibility to open their individual article on condition of the payment of a price similarly than in a gold OA journal. The price level in the hybrid OA is typically around 3,000 USD, which many authors and their institutions perceive as high (Tenopir et al., 2017).

Hybrid journals are a risk free transition path towards full OA, in contrast to starting new full OA journals or converting ones, since the subscription revenue remains (Prosser, 2003). Thus, since Springer announced in 2004 the hybrid option "Open Choice" for their full portfolio of subscription journals, most big publishers have adopted similar modalities and the number of journals offering the hybrid possibility has increased in recent years.

The vast majority of subscription journals from the leading scholarly publishers are nowadays hybrid. The number of journals offering the hybrid option has increased from around 2,000 in the year 2009 to almost 10,000 in the year 2016, and the number of individual articles in the same period has grown from an estimated 8,000 in the year 2009 to 45,000 in the year 2016 (Björk, 2017).

Since Lawrence proposed in 2001 the OA citation advantage, this postulate has been discussed in depth without an agreement being reached (Davis et al., 2008). Furthermore, some authors are critical about the causal link between OA and higher citations, stating that the benefits of OA are uncertain and vary among different fields (Davis & Walters, 2011).

In this paper, as novel contribution, we take a journal-level approach to assessing the OA citation advantage, while many others take a paper-level approach. This is because research articles in both publication modalities in the same hybrid journal and publication year, are quite similar in discipline and with a priori the same citation potential.

Thus, based on citation data from the Scopus database, we provide longitudinal estimations of cites per article in both publication modalities in hybrid journals. Moreover, we answer the following questions:

1. Are hybrid OA research articles more highly cited than their paywalled counterparts?

2. How does this difference vary according to field and time?

**Theoretical framework on open access citation advantage**

Many researchers, starting with Lawrence (2001), have found that OA articles tend to have more citations than pay-for-access articles. This OA citation advantage has been observed in a variety of academic fields including computer science (Lawrence, 2001), philosophy, political science, electrical & electronic engineering, and mathematics (Antelman, 2004), physics (Harnad et al., 2004), biology and chemistry (Eysenbach, 2006), as well as civil engineering (Koler-Povh et al., 2014).

However, this postulate has been discussed in the literature in depth without an agreement being reached (Davis et al., 2008; Dorta-González & Santana-Jiménez, 2018; Norris et al., 2008; Joint, 2009; Gargouri et al., 2010; González-Betancor & Dorta-González, 2019; Wang et al., 2015). Furthermore, some authors are critical about the causal link between OA and higher citations, stating that the benefits of OA are uncertain and vary among different fields (Craig et al., 2007; Davis & Walters, 2011).

Kurtz et al. (2005), and later other authors (Craig et al., 2007; Moed, 2007; Davis et al., 2008), set out three postulates supporting the existence of a correlation between OA and increased citations. (1) OA articles are easier to obtain, and therefore easier to read and cite (*Open Access postulate*). (2) OA articles tend to be available online prior to their publication and therefore begin accumulating citations earlier than pay-for-access articles (*Early View postulate*). And (3) more prominent authors are more likely to provide OA to their articles, and authors are more likely to provide OA to their highest quality articles (*Selection Bias postulate*). Moreover, these authors conclude that early view and selection bias effects are the main factors behind this correlation.

Gaule & Maystre (2011) and Niyazov et al. (2016) found evidence of selection bias in OA, but still estimated a statistically significant citation advantage even after controlling for that bias. Regardless of the validity or generality of their conclusions, these studies establish that any analysis must take into account the effect of time (citation time window) and selection bias.

At journal level, Gumpenberger et al. (2013) showed that the impact factor of gold OA journals was increasing, and that one-third of newly launched OA journals were indexed in the Journal Citation Reports (JCR) after three years. However, Björk and Solomon (2012) argued that the economic model is not related to journal impact. This result has been confirmed by Solomon et al. (2013), concluding that articles are cited at a similar rate regardless of the distribution model.

The OA citation advantage is not universally supported. Many studies have been criticised on methodological grounds (Davis & Walters, 2011), and a research using the randomized-control trial method failed to find evidence of an OA citation advantage (Davis, 2011).

However, recent studies using robust methods have observed an OA citation advantage. McCabe & Snyder (2014) used a complex statistical model to remove author bias and reported a small but meaningful 8% OA citation advantage. Archambault et al. (2014) in a massive sample of over one million articles and using field-normalized citation rates, described a 40% OA citation advantage. Ottaviani (2016) reported a 19% OA citation advantage excluding the author self-selection bias and beyond the first years after publication.

In a recent study, Piwowar et al. (2018) used three samples, each of 100,000 articles, to study OA in three populations: all journal articles assigned a DOI, recent journal articles indexed in Web of Science, and articles viewed by users of the open-source browser extension Unpaywall. They estimated that at least 28% of the scholarly literature is OA, and that this proportion is growing mainly in gold and hybrid journals. Accounting for age and discipline, they observed OA articles receive 18% more citations than average, an effect driven primarily by green and hybrid OA.

**Methodology**

Since the end of 2020, Scopus has new Open Access filters providing information on the type of open access per article. With this new classification system, users can now filter their results or use specific OA tags, i.e. gold, hybrid gold, green, and bronze (delayed).

The source of OA information in Scopus is Unpaywall, an open-source browser extension that lets users find OA articles from publishers and repositories (run by OurResearch, a non-profit organization).

In this study, four subject areas in the Scopus database, one in each branch of knowledge, are considered: Arts & Humanities; Economics, Econometrics & Finance; Medicine; and Physics & Astronomy.

We decided a priori to take four subject areas. This number was set so that both figures and tables could be displayed in the paper. The subject areas were selected based on the previous experience of the authors and trying to cover fields as diverse as possible.

For each of these subject areas, the "research articles" in the year 2017 from 50 hybrid journals, and the citations received by such research articles in the period 2017-2020, were downloaded from the Scopus database (April 8, 2021).

Only 2017 was taken as the year of publication (census) in order to have a citation window of at least three full years for all documents (a full window of three years plus the time elapsed during the year of publication). Note that in most areas the maximum of the distribution of citations is reached before the third year from its publication. Articles published at the beginning of 2017 accumulate their citations for almost four years, while those published at the end of 2017 accumulate their citations for just over three years. This consideration has no consequences on the results obtained since the publication under the hybrid open access modality is distributed uniformly among all the issues of the same year.

The 200 journals were randomly selected from those with share of OA papers in 2017 higher than some minimal value: 5% in Medicine, 4% in Arts & Humanities, 2% in Physics & Astronomy, and 2% in Economics, Econometrics & Finance. Said threshold was set based on the prevalence of the OA modality in each subject area, so that this percentage is higher in areas where the OA modality in hybrid journals is more widespread. This information is detailed in the dataset in Annex A.

A total of 2,020,793 "research articles" were published in the Scopus database in 2017, of which 69,093 were in hybrid journals under the OA modality (3.4%). During that same year, the selected four subject areas published 874,556 research articles, of which 33,796 were in hybrid journals with OA modality (3.9%).

The distribution by subject areas is show in Table 1. The hybrid OA prevalence is 4.6% in Medicine, 3.7% in Arts & Humanities, 2.7% in Physics & Astronomy, and 2.5% in Economics, Econometrics & Finance. The four subject areas represents 43.3% of the database in 2017 by including the largest (Medicine) and the fourth largest (Physics & Astronomy) subject areas. Moreover, the OA articles in hybrid journals in the four subject areas represent 48.9% of the database by including also the largest (Medicine) and the fourth largest (Physics & Astronomy) subject areas in hybrid OA articles.

[Table 1 about here]

In the sample, the 62,608 research articles from 200 hybrid journals were analyzed. Of these, 8043 research articles were published under the OA modality. This represents 23.8% of the total OA research articles published in hybrid journals in the subject areas considered (33,796). This information disaggregate by subject areas is show in Table 2. The areas that are overrepresented in the sample in relation to the OA, in relative terms, are Economics, Econometrics & Finance (49.3%) and Arts & Humanities (40.8%). However, in absolute terms, the total number of OA articles included in these two areas are lower than in Medicine and Physics & Astronomy, due to the larger size of the journals in the latter.

[Table 2 about here]

**Results**

*Cites per article in hybrid journals by modality*

About the correlation between variables (Table 3), as expected, the size of the journal does not correlate with any other variable. The OA prevalence in hybrid journals, this is the proportion of research articles under the OA modality, does not correlate with the position of the journal in the citation ranking (best CiteScore percentile). As a particular case, it does weakly and negatively in Arts & Humanities (-0.69), that is, the best-positioned journals in the citation ranking have a lower proportion of OA articles. This is due to some highly prestigious journals that are still in the initial stages of the hybrid publication model.

[Table 3 about here]

The OA prevalence either does not correlate with cites per article in the hybrid modalities. However, the position of the journal in the citation ranking (percentile) correlates weakly with cites per article in both hybrid modalities.

Note the only two variables that present high correlation, above 0.81 in three subject areas, are cites per article according to modality. That is, the higher cites per article in one modality, the greater in the other. Medicine highlight with a very high correlation (0.97). The exception is Physics & Astronomy, where the correlation reduces to 0.49.

As previously commented, there is a strong and positive linear correlation for cites per article in both hybrid modalities (see Figure 1). The coefficient of determination is generally high, with the exception of Physics & Astronomy. The hybrid journals with the greatest impact in one modality are also in the other. The bisector of the square, that is, the imaginary line that begins in the lower left corner and ends in the upper right corner of the square, separates the citation advantage for each modality. The bubbles below the bisector correspond to hybrid journals with citation advantage for the OA modality. Similarly, the bubbles above the bisector correspond to hybrid journals with citation advantage in the paywalled modality (citation disadvantage for the OA). Note in all the areas there is a majority of journals below the bisector, where the citation advantage corresponds to the OA hybrid modality. In fact, the regression line falls below the bisector in all cases, that is, the OA citation advantage in hybrid journals is observed even in the least squares estimate.

[Figure 1 about here]

In relation to the OA prevalence, this is the proportion of articles in the OA modality of the hybrid journal, indicated through the size of the bubble in Figure 1, there is a tendency for big bubbles to gravitate around the origin. This is especially evident for

Humanities and Physics. This means that hybrid journals with higher proportion of OA papers are usually cited less, which is in accordance with mostly negative correlation coefficients for these indicators in Table 3.

The box diagram for the average of cites per article in hybrid journals, according to modality and year of citation, is show in Figure 2. In all subject areas and each citation year, cites per article for those in the OA modality are clearly higher than the citations in the paywalled modality. These average citations for the OA modality are higher both in mean (indicated with the x symbol) and in quartiles of the distribution (box and whisker). Note that the mean of the distribution is considerably larger than the median. This is because the distribution is asymmetric with a long tail on the right.

[Figure 2 about here]

The increase in the number of citations over time relates to the shape of the citation distribution in each subject area. Thus, for example, in Physics & Astronomy the maximum of the distribution reaches in the third year. Beyond this logical increase in the number of citations over years, no clear time effect observes in Figure 2.

*Open Access citation advantage in hybrid journals*

The OA citation advantage (disadvantage when it is negative) for a journal in a particular year, is defined in relation to the sign of the subtraction as follows. If cites per OA article is greater than cites per paywalled article, then the OA citation advantage is:

(Cites per OA - Cites per Paywalled) / Cites per Paywalled.

However, if cites per OA article is less than cites per paywalled article, then the OA citation advantage (disadvantage because de result is negative) is:

(Cites per OA - Cites per Paywalled) / Cites per OA.

The OA citation advantage in relation to the journal percentile shows in Figure 3. There are differences in OA citation advantages between fields. For example, in Medicine there are few journals with a citation disadvantage for the OA, and in most cases the citation advantage is in the range 0–100%. However, in Economics, Econometrics & Finance the differences among journals are much greater and a big number of cases fall into the range from -100% to 200%. Note the only two highly disadvantaged journals have medium percentiles. A more detailed analysis will follow.

[Figure 3 about here]

Figure 4 shows the OA citation advantage by subject areas, with and without outliers. Note the citation advantage of the OA modality in hybrid journals is clear for all subject areas. The data distribution, represented by the box and whiskers, displaces toward the positive part of the vertical axis. The median of the distribution (the inner line that divides the box into two parts) is in the range 25–50%, while the mean is in 40–60%. There is a citation advantage in more than 75% of the journals. Thus, the 25th percentile (the bottom line of the box) is located close to 0% in the worst case (Economics, Econometrics & Finance). Furthermore, the OA citation advantage is consistent across fields (Figure 4) and held in time (Figure 5).

[Figures 4 and 5 about here]

There is OA citation advantage in 80% of hybrid journals (Table 4). In the remaining 20% there are OA citation disadvantage or there are no differences. The results are relatively stable both across fields and over time. The subject areas where the number of journals with OA citation advantage is higher are Medicine (88%) and Arts & Humanities (82%).

[Table 4 about here]

The average of the OA citation advantage (Table 5) increases with time in the area where the OA prevalence is highest (Medicine), but has a U-shape in the area where the OA prevalence is lowest (Economics, Econometrics & Finance).

[Table 5 about here]

For the aggregate citations in 2017-2020, the average OA citation advantage varies in the range 41.4–62.4%, with a mean for the aggregate areas of 50.3%. The highest average reaches in Arts & Humanities, while the lowest observes in Economics, Econometrics & Finance.

[Table 6 about here]

The outliers observed in the data distribution can skew the mean. However, half of the journals have OA citation advantage above the median of the distribution (and the other half below). Thus, the median (Table 6) is more robust measure of central tendency than the mean for data with such a high variance. The median OA citation advantage in 2017-2020 varies among fields in the range 26.9–49.4%, being 36.8% its value for the aggregate areas. The highest median reaches in Arts & Humanities, while the lowest observes in Medicine.

Thus, we can conclude that the citation advantage of the OA modality in hybrid journals, in relation to the paywalled modality, is around 50.3% in average for the 200 journals and four years in the sample, and higher than 36.8% in half of the journals. Moreover, this OA citation advantage held in time. Finally, the highest OA citation advantage is observed in Arts & Humanities.

**Conclusions**

Access to academic literature is a current debate in the research community. Research funders are increasingly mandating OA dissemination while, at the same time, the growth in costs have led more and more university libraries to cancel some subscriptions.

In this paper, the "research articles" in the year 2017 from 200 hybrid journals in four subject areas, and the citations received by such articles in the period 2017-2020 in the Scopus database, were analyzed. The journals were randomly selected from those with share of OA papers higher than some minimal value. More than 60 thousand research articles were analyzed in the sample, of which 24% under the OA modality.

Interestingly, we found that in general, the citations per article in both hybrid modalities strongly correlate. The hybrid journals with the greatest impact in one modality are also in the other. The evidence for this result is weaker in the field of Physics. However, there is no correlation between the OA prevalence, this is the proportion of articles in the OA modality of the hybrid journal, and cites per article in any of the hybrid modalities.

We found that there is OA citation advantage in 80% of hybrid journals. This result is strong both across fields and over time. The number of journals with OA citation advantage is higher in Medicine (88%) and Humanities (82%).

We found that the average OA citation advantage increases with time in the field where the OA prevalence is highest (Medicine), but has a U-shape in the field with lowest OA prevalence (Economics). The average OA citation advantage in 2017-2020 varies among fields in the range 41–62%, with an aggregate mean of 50%. The highest average is obtained in Humanities, while the lowest is observed in Economics.

The median OA citation advantage in 2017-2020 varies in the range 27–49% according to fields, being 37% its value for the aggregate fields. The highest median is observed again in Humanities, while the lowest is obtained in Medicine.

Thus, we can conclude that the citation advantage of the OA modality in hybrid journals, in relation to the paywalled modality, is around 50% in average for the 200 journals and four years in the sample, and higher than 37% in half of the journals. Moreover, the OA citation advantage is consistent across fields and held in time. Finally, the OA citation advantage is higher in Humanities than in Science and Social Science.

There are some considerations in this regard. Some journals in the random sample have been cataloged by the Scopus database as Humanities, but are actually at the intersection with other areas. Notice that there are journals assigned to two different subject categories from two different areas. Indeed, these journals that employ scientific methods with applications to the Humanities receive more citations than pure humanistic journals. Therefore, the results obtained for this area must be taken with caution.

On the reliability of the data source, Unpaywall is indirectly used (through Scopus) to determine the publication modality in hybrid journals. Notice that Unpaywall is based on algorithms and not on indexing. This is the reason why, regardless of the discipline, the OA finder Unpaywall does not locate as many OA versions of journal articles as manual searches (Piwowar et al., 2018; Sergiadis, 2019).

Our study refines previous results by comparing documents more similar to each other, both in discipline and citation potential. Some of the citation advantage in the open access modality is likely due to more access allows more people to read and hence cite articles they otherwise would not. However, causation is difficult to establish and there are many possible bias. Several factors can affect the observed differences in citation rates. Funder mandates can be one of them. Funders are likely to have OA requirement, and well-funded studies are more likely to receive more citations than poorly funded studies (Aagaard et al., 2020).

Another discussed factor is the selection bias postulate (Craig et al., 2007), which suggests that authors choose only their most impactful studies to be open access. Selection bias can occur in both paid open access journals (gold OA) and hybrid journals. This is due to researchers who have financial resources to publish their results prioritize the publication in open access those papers that they consider may have a greater impact. The current study does not examine the cause of the observed citation advantage, but does find that it exists in a very large sample.

Table 1. Description of the subject areas in the study

| | Research Articles in 2017 | | | | |
|---|---|---|---|---|---|
| **Subject Area** | **OA Hybrid** | **%** | **Other modalities *** | **%** | **Total** |
| Arts & Humanities | 2,821 | 3.7% | 74,458 | 96.3% | 77,279 |
| Economics, Econometrics & Finance | 1,097 | 2.5% | 43,376 | 97.5% | 44,473 |
| Medicine | 23,243 | 4.6% | 485,260 | 95.4% | 508,503 |
| Physics & Astronomy | 6,635 | 2.7% | 237,666 | 97.3% | 244,301 |
| **Aggregate Areas** | **33,796** | **3.9%** | **840,760** | **96.1%** | **874,556** |
| Scopus database | 69,093 | 3.4% | 1,951,700 | 96.6% | 2,020,793 |
| % | 48.9% | | 43.1% | | 43.3% |
| *Paywalled modality in hybrid journals, paywalled journals and OA journals | | | | | |

Table 2. Representativeness of the sample

| | Research Articles in Hybrid Journals in 2017 | | | |
|---|---|---|---|---|
| | **Sample** | | **Population** | **Sample %** |
| **Subject Area** | **OA Hybrid** | **Paywalled** | **OA Hybrid** | **OA Hybrid** |
| Arts & Humanities | 1,151 | 5,759 | 2,821 | 40.8% |
| Economics, Econometrics & Finance | 541 | 5,411 | 1,097 | 49.3% |
| Medicine | 4,381 | 15,772 | 23,243 | 18.8% |
| Physics & Astronomy | 1,970 | 27,623 | 6,635 | 29.7% |
| **Total** | **8,043** | **54,565** | **33,796** | **23.8%** |

Table 3. Pearson's linear correlation coefficient

|  | Best CiteScore Percentile 2017 | Research Articles 2017 | OA Prevalence | OA Cites per Article | Paywalled Cites per Article |
|---|---|---|---|---|---|
| **Arts & Humanities** | | | | | |
| Best CiteScore Percentile 2017 | 1.00 | 0.03 | -0.69 | 0.50 | 0.57 |
| Research Articles 2017 | 0.03 | 1.00 | -0.21 | 0.30 | 0.42 |
| OA Prevalence | -0.69 | -0.21 | 1.00 | -0.39 | -0.40 |
| OA Cites per Article | 0.50 | 0.30 | -0.39 | 1.00 | 0.81 |
| Paywalled Cites per Article | 0.57 | 0.42 | -0.40 | 0.81 | 1.00 |
| **Economics, Econometrics & Finance** | | | | | |
| Best CiteScore Percentile 2017 | 1.00 | -0.16 | 0.07 | 0.60 | 0.57 |
| Research Articles 2017 | -0.16 | 1.00 | -0.52 | 0.13 | 0.11 |
| OA Prevalence | 0.07 | -0.52 | 1.00 | -0.14 | -0.12 |
| OA Cites per Article | 0.60 | 0.13 | -0.14 | 1.00 | 0.85 |
| Paywalled Cites per Article | 0.57 | 0.11 | -0.12 | 0.85 | 1.00 |
| **Medicine** | | | | | |
| Best CiteScore Percentile 2017 | 1.00 | -0.48 | -0.22 | 0.29 | 0.34 |
| Research Articles 2017 | -0.48 | 1.00 | -0.14 | -0.18 | -0.20 |
| OA Prevalence | -0.22 | -0.14 | 1.00 | 0.01 | -0.02 |
| OA Cites per Article | 0.29 | -0.18 | 0.01 | 1.00 | 0.97 |
| Paywalled Cites per Article | 0.34 | -0.20 | -0.02 | 0.97 | 1.00 |
| **Physics & Astronomy** | | | | | |
| Best CiteScore Percentile 2017 | 1.00 | 0.12 | 0.00 | 0.33 | 0.52 |
| Research Articles 2017 | 0.12 | 1.00 | -0.35 | 0.03 | 0.15 |
| OA Prevalence | 0.00 | -0.35 | 1.00 | -0.17 | -0.17 |
| OA Cites per Article | 0.33 | 0.03 | -0.17 | 1.00 | 0.49 |
| Paywalled Cites per Article | 0.52 | 0.15 | -0.17 | 0.49 | 1.00 |

Note: (a) The OA prevalence is the proportion of articles in the OA modality of the hybrid journal. (b) We use the term 'Best percentile' because a journal may be assigned to several subject fields and have different percentiles in each of them.

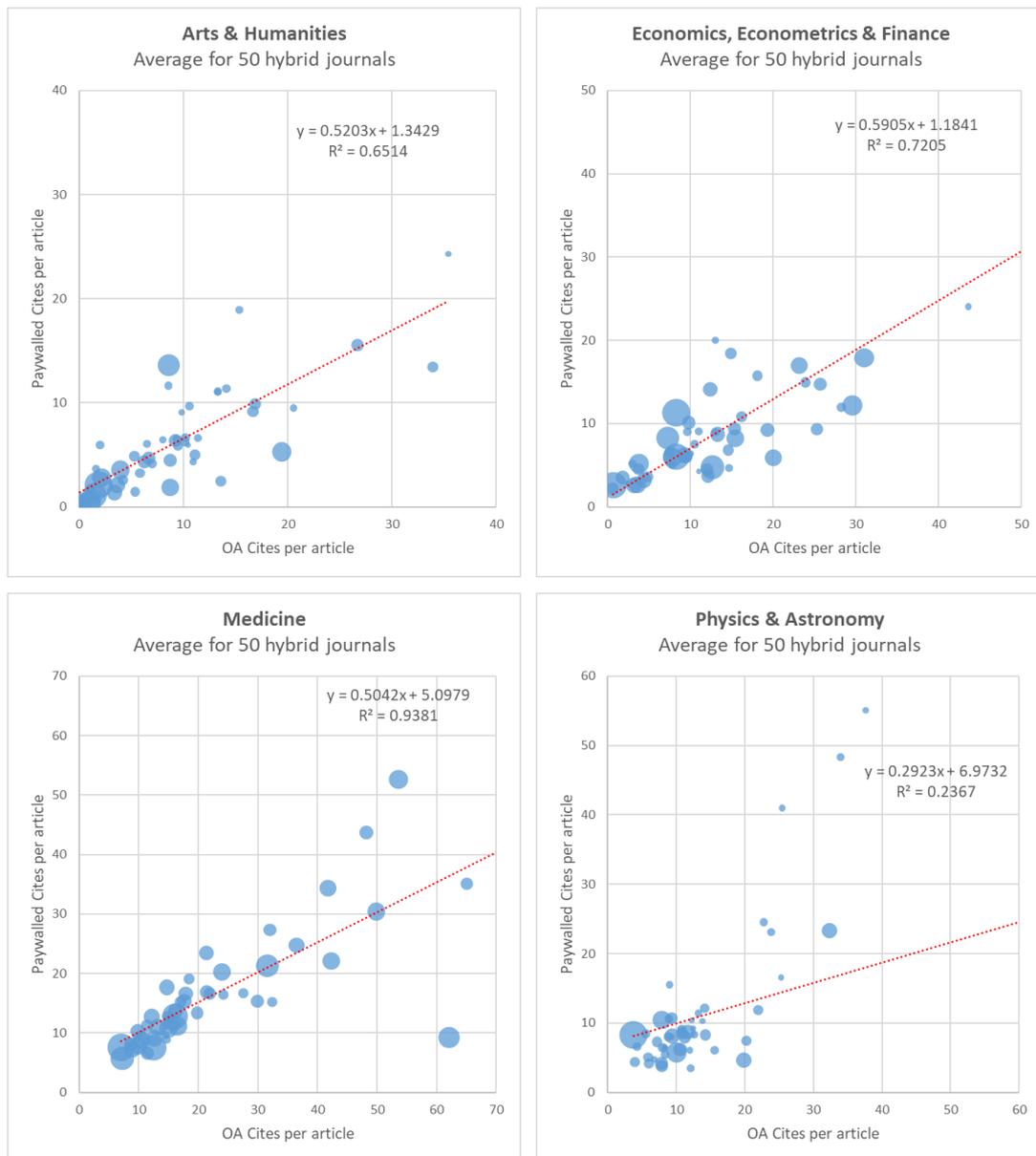

Figure 1. Scatter plot for cites per article in both hybrid modalities. Average across all citation years for the 200 hybrid journals in the sample. Bubble size proportional to OA prevalence.

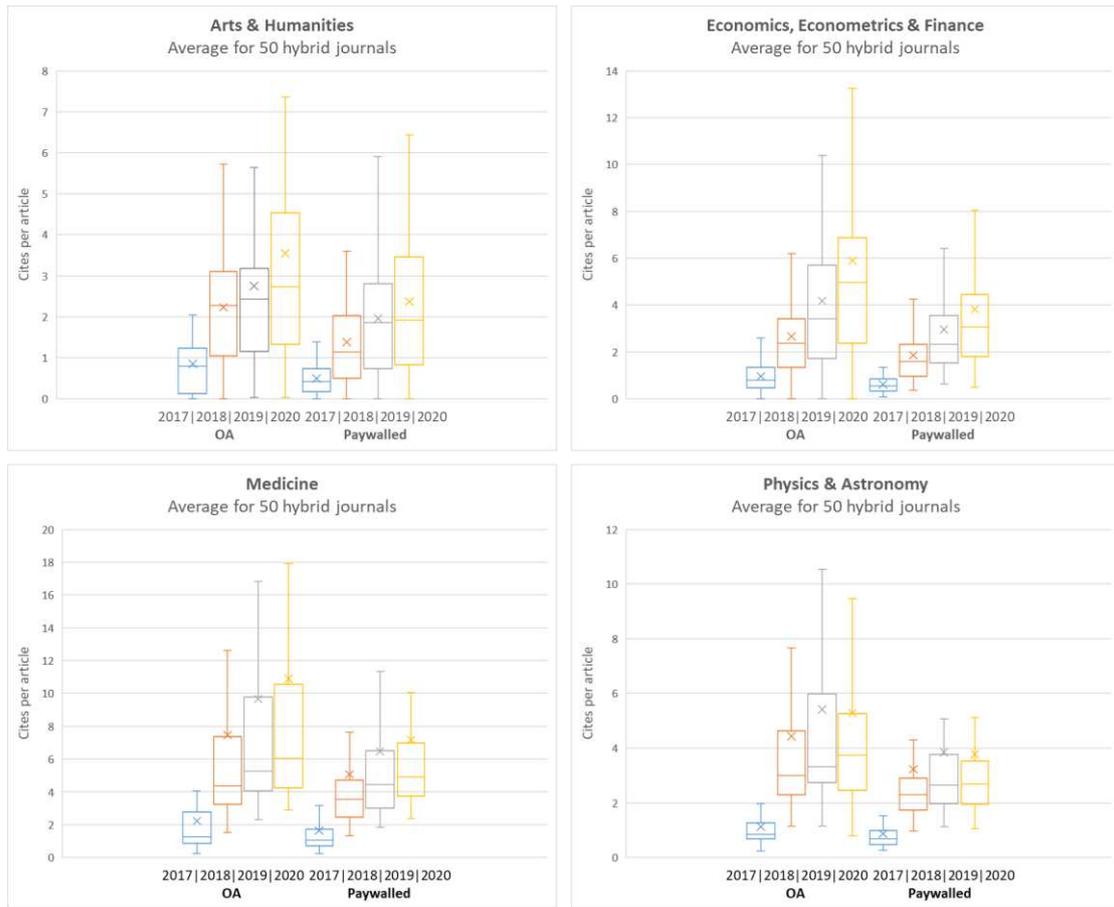

Figure 2. Box and whisker plot (without outliers) for the distribution of cites per article by hybrid modality and year of citation. Average in the citation years for the 200 hybrid journals in the sample.

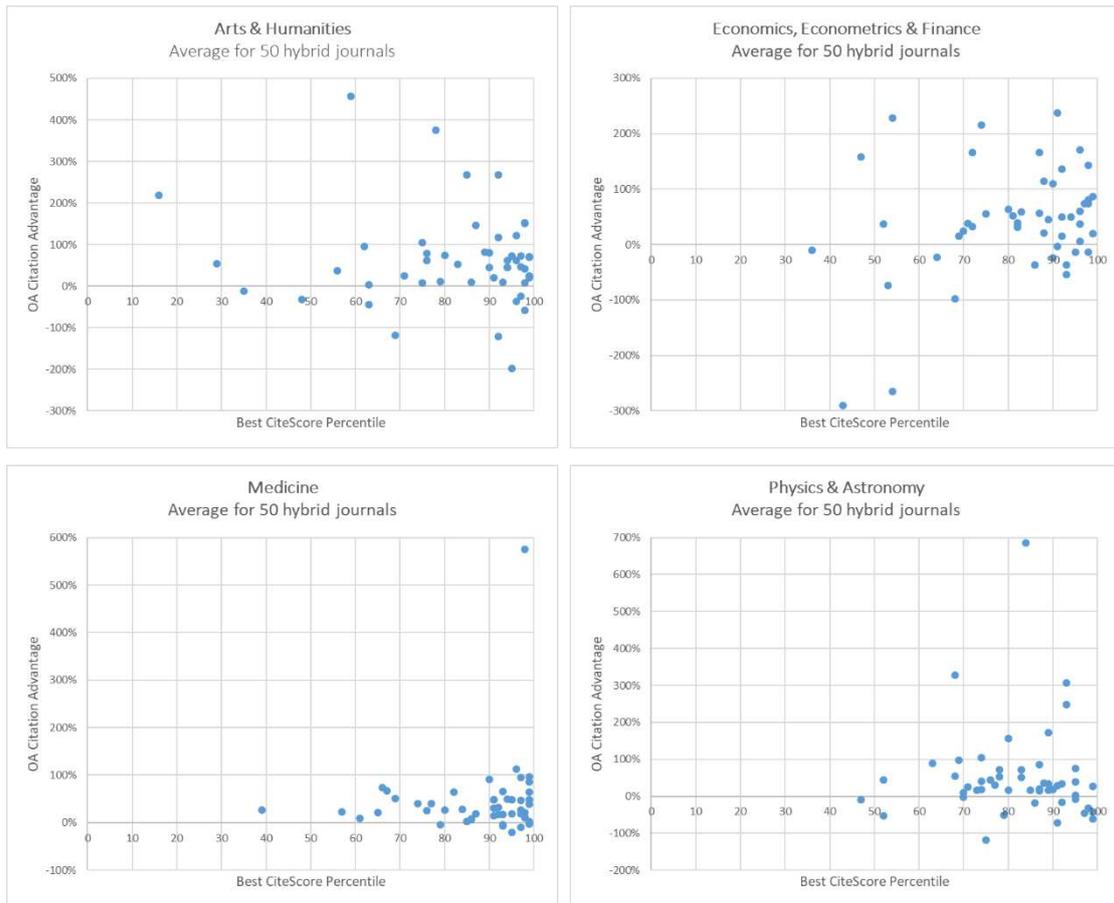

Figure 3. OA citation advantage in relation to the best CiteScore percentile. We use the term 'Best percentile' because a journal may be assigned to several subject fields and have different percentiles in each of them.

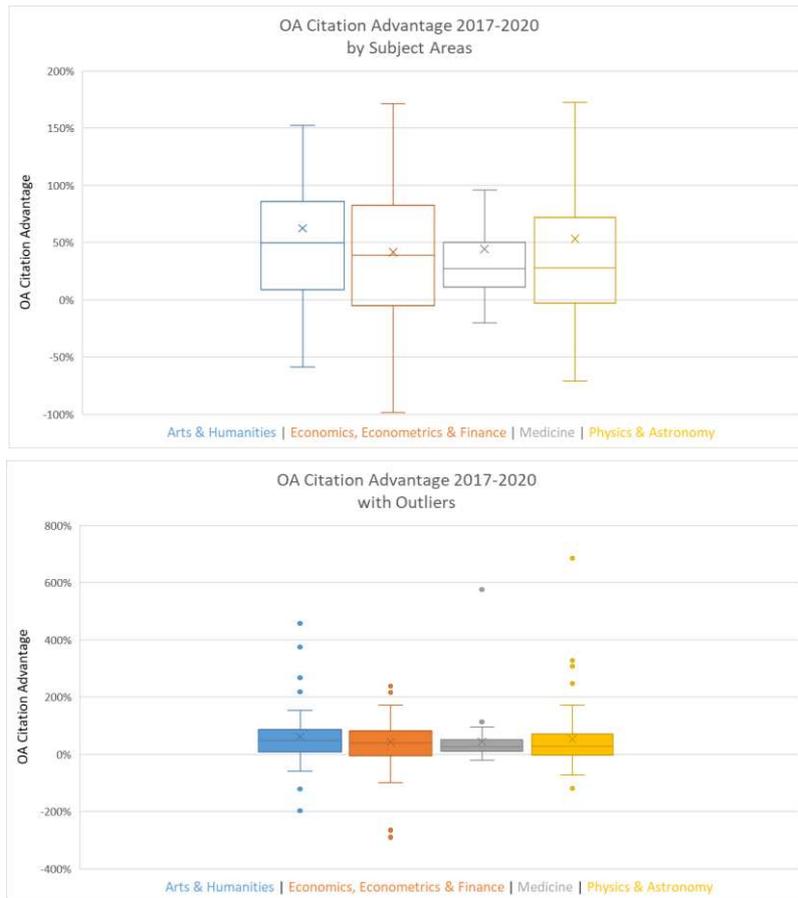

Figure 4. OA citation advantage by subject areas

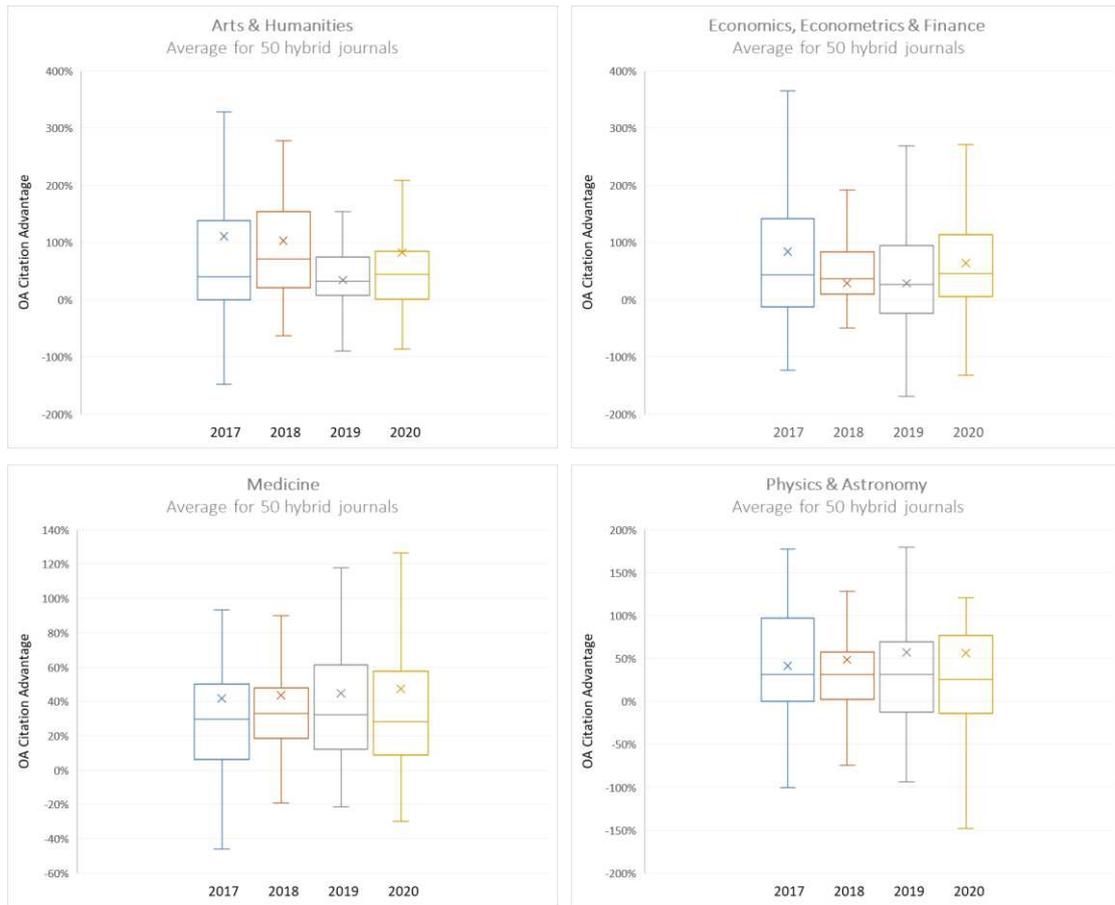

Figure 5. OA citation advantage along time

Table 4. Number of journals with OA citation advantage

| Subject Area | Number of Journals with OA Citation Advantage | | | | | | | | | |
| --- | --- | --- | --- | --- | --- | --- | --- | --- | --- | --- |
| | 2017 | | 2018 | | 2019 | | 2020 | | 2017-2020 | |
| Arts & Humanities | 33 | 66% | 41 | 82% | 40 | 80% | 39 | 78% | 41 | 82% |
| Economics, Econometrics & Finance | 35 | 70% | 38 | 76% | 34 | 68% | 40 | 80% | 37 | 74% |
| Medicine | 39 | 78% | 43 | 86% | 44 | 88% | 40 | 80% | 44 | 88% |
| Physics & Astronomy | 38 | 76% | 38 | 76% | 35 | 70% | 36 | 72% | 37 | 74% |
| **Aggregate areas** | **145** | **73%** | **160** | **80%** | **153** | **77%** | **155** | **78%** | **159** | **80%** |

Table 5. Mean of the OA citation advantage

| Subject Area | Mean OA Citation Advantage | | | | |
| --- | --- | --- | --- | --- | --- |
| | 2017 | 2018 | 2019 | 2020 | 2017-2020 |
| Arts & Humanities | 110.6% | 102.8% | 34.4% | 82.5% | 62.4% |
| Economics, Econometrics & Finance | 83.9% | 28.4% | 28.8% | 64.1% | 41.4% |
| Medicine | 41.7% | 43.5% | 44.7% | 47.4% | 44.3% |
| Physics & Astronomy | 41.4% | 48.6% | 57.0% | 56.3% | 53.1% |
| **Aggregate areas** | | | | | **50.3%** |

Table 6. Median of the OA citation advantage

| Subject Area | Median OA Citation Advantage | | | | |
| --- | --- | --- | --- | --- | --- |
| | 2017 | 2018 | 2019 | 2020 | 2017-2020 |
| Arts & Humanities | 38.9% | 62.9% | 31.8% | 44.1% | 49.4% |
| Economics, Econometrics & Finance | 41.8% | 37.9% | 24.2% | 45.8% | 39.0% |
| Medicine | 29.2% | 32.1% | 31.5% | 28.0% | 26.9% |
| Physics & Astronomy | 30.5% | 30.0% | 30.7% | 24.8% | 27.9% |
| **Aggregate areas** | | | | | **36.8%** |

# ANNEX A. Dataset

| Journal | Best CiteScore Percentile 2017 | Modality | Research Articles 2017 | % | Cites 2017 | Cites 2018 | Cites 2019 | Cites 2020 | Total Cites | Cites per Article | OA Citation Advantage |
|---|---|---|---|---|---|---|---|---|---|---|---|
| **Arts and Humanities** | | | | | | | | | | | |
| Anales de Literatura Hispanoamericana | 29 | Paywalled | 20 | 43.5% | 0 | 1 | 0 | 0 | 1 | 0.05 | **53.8%** |
| | | OA | 26 | 56.5% | 0 | 0 | 1 | 1 | 2 | 0.08 | |
| Archaeological and Anthropological Sciences | 94 | Paywalled | 102 | 90.3% | 86 | 157 | 202 | 217 | 662 | 6.49 | **44.3%** |
| | | OA | 11 | 9.7% | 11 | 27 | 26 | 39 | 103 | 9.36 | |
| Archaeology Ethnology and Anthropology of Eurasia | 56 | Paywalled | 12 | 27.9% | 0 | 2 | 5 | 6 | 13 | 1.08 | **37.0%** |
| | | OA | 31 | 72.1% | 2 | 10 | 14 | 20 | 46 | 1.48 | |
| Archaeometry | 99 | Paywalled | 64 | 88.9% | 35 | 62 | 85 | 83 | 265 | 4.14 | **69.1%** |
| | | OA | 8 | 11.1% | 3 | 21 | 14 | 18 | 56 | 7.00 | |
| Archives of Design Research | 16 | Paywalled | 14 | 29.2% | 0 | 1 | 2 | 1 | 4 | 0.29 | **219.1%** |
| | | OA | 34 | 70.8% | 1 | 8 | 12 | 10 | 31 | 0.91 | |
| Archives of Sexual Behavior | 91 | Paywalled | 197 | 94.7% | 274 | 484 | 692 | 727 | 2177 | 11.05 | **20.1%** |
| | | OA | 11 | 5.3% | 10 | 30 | 45 | 61 | 146 | 13.27 | |
| Artnodes | 69 | Paywalled | 11 | 47.8% | 0 | 0 | 3 | 1 | 4 | 0.36 | **-118.2%** |
| | | OA | 12 | 52.2% | 0 | 0 | 1 | 1 | 2 | 0.17 | |
| Attention Perception and Psychophysics | 94 | Paywalled | 177 | 88.9% | 61 | 261 | 350 | 363 | 1035 | 5.85 | **61.7%** |
| | | OA | 22 | 11.1% | 22 | 52 | 69 | 65 | 208 | 9.45 | |
| Behavior Research Methods | 95 | Paywalled | 132 | 82.0% | 132 | 416 | 654 | 850 | 2052 | 15.55 | **71.9%** |
| | | OA | 29 | 18.0% | 52 | 163 | 225 | 335 | 775 | 26.72 | |
| Bioethics | 96 | Paywalled | 59 | 89.4% | 39 | 109 | 111 | 112 | 371 | 6.29 | **61.3%** |
| | | OA | 7 | 10.6% | 11 | 21 | 23 | 16 | 71 | 10.14 | |
| Biology and Philosophy | 96 | Paywalled | 48 | 85.7% | 21 | 51 | 68 | 100 | 240 | 5.00 | **122.5%** |
| | | OA | 8 | 14.3% | 15 | 28 | 25 | 21 | 89 | 11.13 | |
| Brain and Cognition | 92 | Paywalled | 96 | 93.2% | 56 | 205 | 302 | 348 | 911 | 9.49 | **116.8%** |
| | | OA | 7 | 6.8% | 26 | 40 | 34 | 44 | 144 | 20.57 | |
| Brain and Language | 98 | Paywalled | 97 | 94.2% | 62 | 241 | 267 | 310 | 880 | 9.07 | **8.4%** |
| | | OA | 6 | 5.8% | 2 | 19 | 18 | 20 | 59 | 9.83 | |
| Cognition | 99 | Paywalled | 222 | 91.0% | 173 | 632 | 822 | 889 | 2516 | 11.33 | **24.3%** |
| | | OA | 22 | 9.0% | 20 | 82 | 95 | 113 | 310 | 14.09 | |
| Cognition and Emotion | 86 | Paywalled | 141 | 91.0% | 151 | 318 | 419 | 480 | 1368 | 9.70 | **9.0%** |
| | | OA | 14 | 9.0% | 19 | 21 | 43 | 65 | 148 | 10.57 | |
| Computers in Human Behavior | 97 | Paywalled | 634 | 95.6% | 715 | 2976 | 5094 | 6609 | 15394 | 24.28 | **45.9%** |
| | | OA | 29 | 4.4% | 39 | 207 | 344 | 437 | 1027 | 35.41 | |
| Contemporary British History | 87 | Paywalled | 19 | 70.4% | 4 | 3 | 9 | 10 | 26 | 1.37 | **146.6%** |
| | | OA | 8 | 29.6% | 5 | 9 | 5 | 8 | 27 | 3.38 | |
| Cultural and Social History | 75 | Paywalled | 30 | 81.1% | 3 | 4 | 4 | 12 | 23 | 0.77 | **105.0%** |
| | | OA | 7 | 18.9% | 0 | 3 | 6 | 2 | 11 | 1.57 | |
| Design Journal | 35 | Paywalled | 45 | 9.9% | 5 | 17 | 34 | 39 | 95 | 2.11 | **-12.3%** |
| | | OA | 410 | 90.1% | 3 | 136 | 259 | 373 | 771 | 1.88 | |
| Ethical Theory and Moral Practice | 76 | Paywalled | 41 | 69.5% | 5 | 17 | 27 | 35 | 84 | 2.05 | **79.0%** |
| | | OA | 18 | 30.5% | 3 | 12 | 21 | 30 | 66 | 3.67 | |
| Evolution and Human Behavior | 96 | Paywalled | 79 | 91.9% | 90 | 195 | 282 | 353 | 920 | 11.65 | **-35.9%** |
| | | OA | 7 | 8.1% | 6 | 17 | 16 | 21 | 60 | 8.57 | |
| Gender Place and Culture | 97 | Paywalled | 97 | 92.4% | 30 | 123 | 212 | 276 | 641 | 6.61 | **72.1%** |
| | | OA | 8 | 7.6% | 1 | 20 | 20 | 50 | 91 | 11.38 | |
| Human Ecology | 93 | Paywalled | 58 | 85.3% | 14 | 56 | 107 | 103 | 280 | 4.83 | **9.8%** |
| | | OA | 10 | 14.7% | 1 | 12 | 13 | 27 | 53 | 5.30 | |
| International Journal of Audiology | 95 | Paywalled | 107 | 89.9% | 44 | 151 | 233 | 209 | 637 | 5.95 | **-197.7%** |
| | | OA | 12 | 10.1% | 6 | 20 | 20 | 20 | 24 | 2.00 | |
| International Journal of Transpersonal Studies | 48 | Paywalled | 11 | 57.9% | 12 | 1 | 10 | 8 | 31 | 2.82 | **-32.6%** |
| | | OA | 8 | 42.1% | 0 | 5 | 8 | 4 | 17 | 2.13 | |
| Journal of Agricultural and Environmental Ethics | 98 | Paywalled | 34 | 79.1% | 6 | 35 | 63 | 56 | 160 | 4.71 | **41.7%** |
| | | OA | 9 | 20.9% | 14 | 11 | 17 | 18 | 60 | 6.67 | |
| Journal of Archaeological Science | 99 | Paywalled | 105 | 91.3% | 107 | 275 | 350 | 429 | 1161 | 11.06 | **20.3%** |
| | | OA | 10 | 8.7% | 17 | 32 | 44 | 40 | 133 | 13.30 | |
| Journal of Archaeological Science Reports | 80 | Paywalled | 379 | 95.9% | 177 | 545 | 737 | 806 | 2265 | 5.98 | **74.6%** |
| | | OA | 16 | 4.1% | 11 | 42 | 49 | 65 | 167 | 10.44 | |
| Journal of Business Ethics | 97 | Paywalled | 300 | 92.6% | 345 | 953 | 1772 | 2610 | 5680 | 18.93 | **-23.5%** |
| | | OA | 24 | 7.4% | 27 | 82 | 108 | 151 | 368 | 15.33 | |
| Journal of Ethnic and Migration Studies | 89 | Paywalled | 124 | 83.8% | 126 | 213 | 338 | 457 | 1134 | 9.15 | **82.2%** |
| | | OA | 24 | 16.2% | 49 | 79 | 103 | 169 | 400 | 16.67 | |
| Journal of Intellectual and Developmental Disability | 62 | Paywalled | 30 | 78.9% | 15 | 31 | 34 | 54 | 134 | 4.47 | **95.9%** |
| | | OA | 8 | 21.1% | 8 | 17 | 23 | 22 | 70 | 8.75 | |
| Journal of Medical Ethics | 90 | Paywalled | 113 | 80.1% | 82 | 200 | 218 | 221 | 721 | 6.38 | **44.4%** |
| | | OA | 28 | 19.9% | 28 | 69 | 70 | 91 | 258 | 9.21 | |
| Journal of Memory and Language | 98 | Paywalled | 88 | 87.1% | 97 | 311 | 346 | 428 | 1182 | 13.43 | **152.6%** |
| | | OA | 13 | 12.9% | 20 | 98 | 128 | 195 | 441 | 33.92 | |
| Journal of Psycholinguistic Research | 76 | Paywalled | 79 | 87.8% | 15 | 42 | 54 | 93 | 204 | 2.58 | **61.9%** |
| | | OA | 11 | 12.2% | 8 | 12 | 13 | 13 | 46 | 4.18 | |
| Journal of Religion and Health | 98 | Paywalled | 163 | 93.7% | 75 | 155 | 215 | 263 | 708 | 4.34 | **151.2%** |
| | | OA | 11 | 6.3% | 7 | 34 | 31 | 48 | 120 | 10.91 | |
| Journal of Southern African Studies | 59 | Paywalled | 59 | 85.5% | 9 | 39 | 50 | 46 | 144 | 2.44 | **457.2%** |
| | | OA | 10 | 14.5% | 35 | 25 | 31 | 45 | 136 | 13.60 | |
| Journal of the Acoustical Society of America | 71 | Paywalled | 805 | 93.2% | 322 | 1364 | 1850 | 1642 | 5178 | 6.43 | **24.6%** |
| | | OA | 59 | 6.8% | 28 | 120 | 167 | 158 | 473 | 8.02 | |
| Journal of World Prehistory | 98 | Paywalled | 5 | 41.7% | 2 | 18 | 20 | 28 | 68 | 13.60 | **-58.7%** |
| | | OA | 7 | 58.3% | 6 | 17 | 16 | 21 | 60 | 8.57 | |
| Law and Philosophy | 78 | Paywalled | 12 | 63.2% | 2 | 12 | 5 | 3 | 22 | 1.83 | **375.3%** |
| | | OA | 7 | 36.8% | 0 | 15 | 22 | 24 | 61 | 8.71 | |
| Medical Humanities | 85 | Paywalled | 81 | 88.0% | 11 | 23 | 35 | 49 | 118 | 1.46 | **268.2%** |
| | | OA | 11 | 12.0% | 11 | 11 | 14 | 23 | 59 | 5.36 | |
| Memory | 75 | Paywalled | 123 | 92.5% | 69 | 172 | 253 | 251 | 745 | 6.06 | **7.3%** |
| | | OA | 10 | 7.5% | 6 | 14 | 27 | 18 | 65 | 6.50 | |
| Memory and Cognition | 83 | Paywalled | 93 | 90.3% | 47 | 149 | 185 | 239 | 620 | 6.67 | **53.0%** |
| | | OA | 10 | 9.7% | 12 | 23 | 23 | 44 | 102 | 10.20 | |
| Neophilologus | 63 | Paywalled | 35 | 81.4% | 1 | 2 | 3 | 11 | 17 | 0.49 | **2.9%** |
| | | OA | 8 | 18.6% | 0 | 0 | 1 | 3 | 4 | 0.50 | |
| Philosophia United States | 63 | Paywalled | 105 | 84.7% | 9 | 27 | 63 | 60 | 159 | 1.51 | **-43.9%** |
| | | OA | 19 | 15.3% | 0 | 3 | 6 | 11 | 20 | 1.05 | |
| Philosophical Explorations | 79 | Paywalled | 23 | 59.0% | 4 | 15 | 28 | 35 | 82 | 3.57 | **10.4%** |
| | | OA | 16 | 41.0% | 2 | 17 | 22 | 22 | 63 | 3.94 | |
| Philosophical Studies | 92 | Paywalled | 162 | 92.0% | 46 | 113 | 192 | 238 | 589 | 3.64 | **-121.3%** |
| | | OA | 14 | 8.0% | 1 | 7 | 1 | 14 | 23 | 1.64 | |
| Philosophy and Technology | 92 | Paywalled | 14 | 56.0% | 6 | 17 | 18 | 33 | 74 | 5.29 | **268.1%** |
| | | OA | 11 | 44.0% | 17 | 41 | 75 | 81 | 214 | 19.45 | |
| Political Geography | 99 | Paywalled | 72 | 83.7% | 39 | 143 | 233 | 295 | 710 | 9.86 | **70.9%** |
| | | OA | 14 | 16.3% | 15 | 51 | 79 | 91 | 236 | 16.86 | |
| Review of Philosophy and Psychology | 94 | Paywalled | 33 | 80.5% | 14 | 27 | 42 | 60 | 143 | 4.33 | **44.2%** |
| | | OA | 8 | 19.5% | 7 | 18 | 9 | 16 | 50 | 6.25 | |
| Synthese | 90 | Paywalled | 209 | 88.2% | 42 | 139 | 203 | 290 | 674 | 3.22 | **80.5%** |
| | | OA | 28 | 11.8% | 19 | 38 | 47 | 59 | 163 | 5.82 | |
| **Economics, Econometrics & Finance** | | | | | | | | | | | |
| Applied Economics | 47 | Paywalled | 410 | 97.9% | 98 | 336 | 561 | 752 | 1747 | 4.26 | **158.2%** |
| | | OA | 9 | 2.1% | 10 | 21 | 19 | 49 | 99 | 11.00 | |

| Journal | | | | | | | | | | |
|---|---|---|---|---|---|---|---|---|---|---|
| Applied Economics Letters | 36 | Paywalled | 307 | 97.5% | 71 | 157 | 286 | 332 | 846 | 2.76 | -10.2% |
| | | OA | 8 | 2.5% | 1 | 7 | 3 | 9 | 20 | 2.50 | |
| Applied Health Economics and Health Policy | 83 | Paywalled | 49 | 89.1% | 21 | 74 | 84 | 114 | 293 | 5.98 | 58.9% |
| | | OA | 6 | 10.9% | 9 | 10 | 22 | 16 | 57 | 9.50 | |
| Computational Economics | 72 | Paywalled | 54 | 91.5% | 19 | 42 | 66 | 68 | 195 | 3.61 | 32.9% |
| | | OA | 5 | 8.5% | 4 | 7 | 5 | 8 | 24 | 4.80 | |
| Ecological Economics | 92 | Paywalled | 266 | 92.4% | 255 | 953 | 1330 | 1654 | 4192 | 15.76 | 14.8% |
| | | OA | 22 | 7.6% | 29 | 96 | 123 | 150 | 398 | 18.09 | |
| Economic Geography | 98 | Paywalled | 15 | 75.0% | 15 | 64 | 89 | 100 | 268 | 17.87 | 73.5% |
| | | OA | 5 | 25.0% | 7 | 24 | 52 | 72 | 155 | 31.00 | |
| Economic Journal | 91 | Paywalled | 105 | 87.5% | 99 | 218 | 323 | 422 | 1062 | 10.11 | -3.2% |
| | | OA | 15 | 12.5% | 9 | 29 | 49 | 60 | 147 | 9.80 | |
| Economic Theory | 68 | Paywalled | 59 | 86.8% | 35 | 44 | 53 | 76 | 208 | 3.53 | -98.3% |
| | | OA | 9 | 13.2% | 7 | 1 | 3 | 5 | 16 | 1.78 | |
| Economics Letters | 53 | Paywalled | 390 | 95.6% | 53 | 356 | 761 | 859 | 2029 | 5.20 | -73.4% |
| | | OA | 18 | 4.4% | 5 | 11 | 14 | 24 | 54 | 3.00 | |
| Economist Netherlands | 43 | Paywalled | 10 | 52.6% | 8 | 4 | 9 | 5 | 26 | 2.60 | -290.0% |
| | | OA | 9 | 47.4% | 4 | 0 | 2 | 0 | 6 | 0.67 | |
| Empirical Economics | 64 | Paywalled | 126 | 90.6% | 27 | 133 | 173 | 241 | 574 | 4.56 | -23.4% |
| | | OA | 13 | 9.4% | 4 | 4 | 14 | 26 | 48 | 3.69 | |
| Energy Economics | 93 | Paywalled | 342 | 96.9% | 333 | 1352 | 2193 | 2966 | 6844 | 20.01 | -53.9% |
| | | OA | 11 | 3.1% | 8 | 31 | 50 | 54 | 143 | 13.00 | |
| Environmental and Resource Economics | 81 | Paywalled | 95 | 84.8% | 80 | 167 | 279 | 303 | 829 | 8.73 | 52.3% |
| | | OA | 17 | 15.2% | 28 | 49 | 62 | 87 | 226 | 13.29 | |
| European Journal of Health Economics | 95 | Paywalled | 53 | 65.4% | 55 | 98 | 130 | 156 | 439 | 8.28 | -13.7% |
| | | OA | 28 | 34.6% | 13 | 58 | 49 | 84 | 204 | 7.29 | |
| Experimental Economics | 94 | Paywalled | 35 | 83.3% | 16 | 56 | 74 | 71 | 217 | 6.20 | 49.8% |
| | | OA | 7 | 16.7% | 7 | 17 | 25 | 16 | 65 | 9.29 | |
| Finance and Stochastics | 87 | Paywalled | 20 | 62.5% | 6 | 28 | 32 | 29 | 95 | 4.75 | 166.7% |
| | | OA | 12 | 37.5% | 4 | 26 | 49 | 73 | 152 | 12.67 | |
| Fiscal Studies | 82 | Paywalled | 18 | 75.0% | 11 | 20 | 29 | 47 | 107 | 5.94 | 31.8% |
| | | OA | 6 | 25.0% | 6 | 6 | 13 | 22 | 47 | 7.83 | |
| Food Policy | 98 | Paywalled | 79 | 72.5% | 49 | 203 | 312 | 398 | 962 | 12.18 | 142.8% |
| | | OA | 30 | 27.5% | 60 | 160 | 269 | 398 | 887 | 29.57 | |
| Forest Policy and Economics | 92 | Paywalled | 148 | 94.3% | 198 | 447 | 518 | 604 | 1767 | 11.94 | 136.4% |
| | | OA | 9 | 5.7% | 29 | 60 | 87 | 78 | 254 | 28.22 | |
| Global Policy | 74 | Paywalled | 126 | 95.5% | 72 | 123 | 172 | 218 | 585 | 4.64 | 215.9% |
| | | OA | 6 | 4.5% | 12 | 19 | 28 | 29 | 88 | 14.67 | |
| Globalizations | 88 | Paywalled | 74 | 91.4% | 73 | 121 | 146 | 163 | 503 | 6.80 | 114.4% |
| | | OA | 7 | 8.6% | 5 | 25 | 28 | 44 | 102 | 14.57 | |
| Internat Environmental Agreements Politics Law and Economics | 91 | Paywalled | 37 | 80.4% | 31 | 52 | 68 | 68 | 219 | 5.92 | 237.9% |
| | | OA | 9 | 19.6% | 20 | 35 | 61 | 64 | 180 | 20.00 | |
| International Journal of Agricultural Sustainability | 80 | Paywalled | 43 | 87.8% | 29 | 83 | 120 | 171 | 403 | 9.37 | 63.6% |
| | | OA | 6 | 12.2% | 2 | 16 | 35 | 39 | 92 | 15.33 | |
| International Journal of Game Theory | 54 | Paywalled | 47 | 90.4% | 9 | 31 | 30 | 33 | 103 | 2.19 | -265.2% |
| | | OA | 5 | 9.6% | 0 | 1 | 0 | 2 | 3 | 0.60 | |
| International Journal of Production Economics | 98 | Paywalled | 281 | 96.6% | 331 | 1284 | 2207 | 2930 | 6752 | 24.03 | 81.5% |
| | | OA | 10 | 3.4% | 7 | 80 | 121 | 228 | 436 | 43.60 | |
| Journal of Common Market Studies | 96 | Paywalled | 91 | 94.8% | 70 | 198 | 256 | 298 | 822 | 9.03 | 6.3% |
| | | OA | 5 | 5.2% | 4 | 15 | 8 | 21 | 48 | 9.60 | |
| Journal of Development Economics | 92 | Paywalled | 65 | 91.5% | 32 | 138 | 213 | 319 | 702 | 10.80 | 49.7% |
| | | OA | 6 | 8.5% | 4 | 17 | 34 | 42 | 97 | 16.17 | |
| Journal of Econometrics | 82 | Paywalled | 119 | 95.2% | 67 | 155 | 269 | 403 | 894 | 7.51 | 39.8% |
| | | OA | 6 | 4.8% | 8 | 10 | 15 | 30 | 63 | 10.50 | |
| Journal of Economic Behavior and Organization | 75 | Paywalled | 221 | 96.5% | 53 | 266 | 463 | 635 | 1417 | 6.41 | 56.0% |
| | | OA | 8 | 3.5% | 4 | 18 | 16 | 42 | 80 | 10.00 | |
| Journal of Economic Inequality | 86 | Paywalled | 14 | 73.7% | 2 | 12 | 30 | 29 | 73 | 5.21 | -37.2% |
| | | OA | 5 | 26.3% | 4 | 2 | 5 | 8 | 19 | 3.80 | |
| Journal of Environmental Economics and Management | 90 | Paywalled | 71 | 91.0% | 85 | 259 | 392 | 571 | 1307 | 18.41 | -23.9% |
| | | OA | 7 | 9.0% | 11 | 27 | 35 | 31 | 104 | 14.86 | |
| Journal of International Economics | 90 | Paywalled | 81 | 87.1% | 30 | 142 | 225 | 349 | 746 | 9.21 | 109.9% |
| | | OA | 12 | 12.9% | 9 | 38 | 71 | 114 | 232 | 19.33 | |
| Journal of International Money and Finance | 88 | Paywalled | 116 | 95.9% | 52 | 187 | 316 | 499 | 1054 | 9.09 | 21.1% |
| | | OA | 5 | 4.1% | 5 | 7 | 16 | 27 | 55 | 11.00 | |
| Journal of Public Economics | 89 | Paywalled | 113 | 95.0% | 47 | 184 | 333 | 449 | 1013 | 8.96 | 45.0% |
| | | OA | 6 | 5.0% | 7 | 15 | 18 | 38 | 78 | 13.00 | |
| Journal of Risk and Uncertainty | 82 | Paywalled | 11 | 52.4% | 1 | 17 | 26 | 22 | 66 | 6.00 | 38.3% |
| | | OA | 10 | 47.6% | 0 | 18 | 27 | 38 | 83 | 8.30 | |
| Journal of the Academy of Marketing Science | 99 | Paywalled | 45 | 90.0% | 125 | 320 | 663 | 1045 | 2153 | 47.84 | 20.0% |
| | | OA | 5 | 10.0% | 13 | 41 | 84 | 149 | 287 | 57.40 | |
| Journal of the European Economic Association | 98 | Paywalled | 31 | 86.1% | 24 | 69 | 144 | 201 | 438 | 14.13 | -13.9% |
| | | OA | 5 | 13.9% | 1 | 16 | 18 | 27 | 62 | 12.40 | |
| Labour Economics | 69 | Paywalled | 78 | 92.9% | 31 | 80 | 150 | 247 | 508 | 6.51 | 15.2% |
| | | OA | 6 | 7.1% | 3 | 7 | 15 | 20 | 45 | 7.50 | |
| Letters in Spatial and Resource Sciences | 52 | Paywalled | 21 | 80.8% | 2 | 9 | 22 | 22 | 55 | 2.62 | 37.5% |
| | | OA | 5 | 19.2% | 3 | 2 | 2 | 11 | 18 | 3.60 | |
| Marine Policy | 96 | Paywalled | 304 | 90.2% | 158 | 689 | 964 | 1022 | 2833 | 9.32 | 171.2% |
| | | OA | 33 | 9.8% | 47 | 204 | 266 | 317 | 834 | 25.27 | |
| Public Choice | 71 | Paywalled | 60 | 84.5% | 14 | 36 | 57 | 83 | 190 | 3.17 | 37.8% |
| | | OA | 11 | 15.5% | 2 | 8 | 18 | 20 | 48 | 4.36 | |
| Quantitative Finance | 87 | Paywalled | 109 | 95.6% | 36 | 123 | 175 | 209 | 543 | 4.98 | 56.6% |
| | | OA | 5 | 4.4% | 4 | 7 | 14 | 14 | 39 | 7.80 | |
| Resources Policy | 96 | Paywalled | 121 | 92.4% | 88 | 369 | 594 | 755 | 1806 | 14.93 | 60.1% |
| | | OA | 10 | 7.6% | 8 | 46 | 94 | 91 | 239 | 23.90 | |
| Review of Income and Wealth | 72 | Paywalled | 56 | 88.9% | 19 | 58 | 91 | 84 | 252 | 4.50 | 166.7% |
| | | OA | 7 | 11.1% | 3 | 15 | 27 | 39 | 84 | 12.00 | |
| Review of International Political Economy | 99 | Paywalled | 27 | 79.4% | 9 | 42 | 73 | 99 | 223 | 8.26 | 86.8% |
| | | OA | 7 | 20.6% | 14 | 20 | 32 | 42 | 108 | 15.43 | |
| Small Business Economics | 96 | Paywalled | 80 | 80.8% | 54 | 200 | 479 | 623 | 1356 | 16.95 | 36.6% |
| | | OA | 19 | 19.2% | 19 | 64 | 141 | 216 | 440 | 23.16 | |
| Social Choice and Welfare | 54 | Paywalled | 64 | 87.7% | 29 | 56 | 79 | 72 | 236 | 3.69 | 228.4% |
| | | OA | 9 | 12.3% | 14 | 23 | 29 | 43 | 109 | 12.11 | |
| Technological and Economic Development of Economy | 93 | Paywalled | 22 | 48.9% | 27 | 63 | 82 | 76 | 248 | 11.27 | -36.5% |
| | | OA | 23 | 51.1% | 24 | 54 | 57 | 55 | 190 | 8.26 | |
| Theory and Decision | 70 | Paywalled | 50 | 83.3% | 6 | 18 | 44 | 61 | 129 | 2.58 | 24.0% |
| | | OA | 10 | 16.7% | 1 | 7 | 10 | 14 | 32 | 3.20 | |
| World Development | 97 | Paywalled | 252 | 88.1% | 203 | 746 | 1180 | 1591 | 3720 | 14.76 | 74.1% |
| | | OA | 34 | 11.9% | 35 | 185 | 266 | 388 | 874 | 25.71 | |
| **Medicine** | | | | | | | | | | | |
| Advances in Therapy | 67 | Paywalled | 51 | 38.6% | 26 | 111 | 113 | 133 | 383 | 7.51 | 66.4% |
| | | OA | 81 | 61.4% | 68 | 254 | 358 | 332 | 1012 | 12.49 | |
| American Journal of Cardiology | 82 | Paywalled | 626 | 94.0% | 339 | 1521 | 1853 | 1857 | 5570 | 8.90 | 64.9% |
| | | OA | 40 | 6.0% | 47 | 158 | 202 | 180 | 587 | 14.68 | |
| American Journal of Preventive Medicine | 97 | Paywalled | 237 | 79.5% | 250 | 745 | 1054 | 1190 | 3239 | 13.67 | 18.4% |
| | | OA | 61 | 20.5% | 56 | 198 | 332 | 401 | 987 | 16.18 | |
| American Journal of Tropical Medicine and Hygiene | 66 | Paywalled | 414 | 83.6% | 156 | 678 | 869 | 1022 | 2725 | 6.58 | 73.3% |
| | | OA | 81 | 16.4% | 59 | 231 | 286 | 348 | 924 | 11.41 | |
| Annals of Oncology | 98 | Paywalled | 244 | 82.4% | 572 | 2345 | 3436 | 4319 | 10672 | 43.74 | 10.1% |
| | | OA | 52 | 17.6% | 183 | 656 | 875 | 791 | 2505 | 48.17 | |
| Annals of the Rheumatic Diseases | 99 | Paywalled | 179 | 66.8% | 917 | 2359 | 3031 | 3113 | 9420 | 52.63 | 1.9% |
| | | OA | 89 | 33.2% | 593 | 1378 | 1364 | 1437 | 4772 | 53.62 | |

| Journal | | Type | | | | | | | | | |
|---|---|---|---|---|---|---|---|---|---|---|---|
| Antimicrobial Agents and Chemotherapy | 92 | Paywalled | 751 | 85.2% | 821 | 3242 | 4134 | 4284 | 12481 | 16.62 | **31.7%** |
| | | OA | 130 | 14.8% | 200 | 787 | 937 | 921 | 2845 | 21.88 | |
| Asian Pacific Journal of Tropical Medicine | 93 | Paywalled | 47 | 30.7% | 14 | 62 | 123 | 155 | 354 | 7.53 | **-6.9%** |
| | | OA | 106 | 69.3% | 26 | 163 | 245 | 313 | 747 | 7.05 | |
| BMJ Online | 98 | Paywalled | 186 | 59.6% | 172 | 439 | 557 | 542 | 1710 | 9.19 | **575.7%** |
| | | OA | 126 | 40.4% | 487 | 1836 | 2410 | 3094 | 7827 | 62.12 | |
| Brain | 98 | Paywalled | 178 | 74.8% | 376 | 1512 | 2020 | 2203 | 6111 | 34.33 | **21.6%** |
| | | OA | 60 | 25.2% | 173 | 642 | 781 | 908 | 2504 | 41.73 | |
| Brain Structure and Function | 93 | Paywalled | 200 | 78.1% | 290 | 676 | 756 | 827 | 2549 | 12.75 | **-4.5%** |
| | | OA | 56 | 21.9% | 67 | 182 | 210 | 224 | 683 | 12.20 | |
| Breast Cancer Research and Treatment | 77 | Paywalled | 298 | 83.0% | 251 | 810 | 1110 | 1220 | 3391 | 11.38 | **39.7%** |
| | | OA | 61 | 17.0% | 50 | 244 | 312 | 364 | 970 | 15.90 | |
| British Journal of Dermatology | 97 | Paywalled | 219 | 84.2% | 473 | 833 | 994 | 1059 | 3359 | 15.34 | **94.8%** |
| | | OA | 41 | 15.8% | 133 | 317 | 358 | 417 | 1225 | 29.88 | |
| Cancer Science | 76 | Paywalled | 103 | 40.4% | 26 | 307 | 456 | 527 | 1316 | 12.78 | **25.6%** |
| | | OA | 152 | 59.6% | 184 | 646 | 765 | 845 | 2440 | 16.05 | |
| Cell Systems | 91 | Paywalled | 51 | 54.8% | 49 | 233 | 367 | 437 | 1086 | 21.29 | **48.3%** |
| | | OA | 42 | 45.2% | 90 | 298 | 434 | 504 | 1326 | 31.57 | |
| Clinical Infectious Diseases | 97 | Paywalled | 427 | 80.4% | 842 | 2640 | 3269 | 3264 | 10015 | 23.45 | **-9.9%** |
| | | OA | 104 | 19.6% | 254 | 540 | 715 | 710 | 2219 | 21.34 | |
| Clinical Therapeutics | 80 | Paywalled | 110 | 72.8% | 98 | 191 | 273 | 294 | 856 | 7.78 | **27.3%** |
| | | OA | 41 | 27.2% | 25 | 96 | 140 | 145 | 406 | 9.90 | |
| Diabetes Obesity and Metabolism | 95 | Paywalled | 145 | 71.8% | 285 | 815 | 911 | 918 | 2929 | 20.20 | **18.9%** |
| | | OA | 57 | 28.2% | 164 | 407 | 401 | 397 | 1369 | 24.02 | |
| European Heart Journal | 99 | Paywalled | 299 | 86.7% | 1035 | 2633 | 3219 | 3599 | 10486 | 35.07 | **85.6%** |
| | | OA | 46 | 13.3% | 302 | 769 | 925 | 998 | 2994 | 65.09 | |
| European Radiology | 91 | Paywalled | 485 | 84.8% | 568 | 1784 | 2014 | 2111 | 6477 | 13.35 | **48.3%** |
| | | OA | 87 | 15.2% | 132 | 448 | 546 | 597 | 1723 | 19.80 | |
| Health Policy and Planning | 87 | Paywalled | 116 | 70.7% | 90 | 200 | 292 | 448 | 1030 | 8.88 | **18.0%** |
| | | OA | 48 | 29.3% | 47 | 108 | 169 | 179 | 503 | 10.48 | |
| Human Brain Mapping | 96 | Paywalled | 377 | 90.4% | 384 | 1402 | 1876 | 2061 | 5723 | 15.18 | **113.1%** |
| | | OA | 40 | 9.6% | 76 | 289 | 434 | 495 | 1294 | 32.35 | |
| Human Molecular Genetics | 91 | Paywalled | 322 | 80.7% | 283 | 1219 | 1592 | 1853 | 4947 | 15.36 | **14.7%** |
| | | OA | 77 | 19.3% | 84 | 345 | 464 | 464 | 1357 | 17.62 | |
| Human Vaccines and Immunotherapeutics | 57 | Paywalled | 211 | 75.6% | 93 | 385 | 485 | 564 | 1527 | 7.24 | **23.1%** |
| | | OA | 68 | 24.4% | 37 | 150 | 202 | 217 | 606 | 8.91 | |
| Infection and Immunity | 85 | Paywalled | 240 | 86.0% | 207 | 687 | 868 | 915 | 2677 | 11.15 | **2.1%** |
| | | OA | 39 | 14.0% | 43 | 98 | 150 | 153 | 444 | 11.38 | |
| International Journal of Cancer | 93 | Paywalled | 430 | 90.0% | 578 | 1880 | 2275 | 2441 | 7174 | 16.68 | **65.2%** |
| | | OA | 48 | 10.0% | 67 | 341 | 420 | 495 | 1323 | 27.56 | |
| International Journal of Cardiology | 69 | Paywalled | 1104 | 94.6% | 1024 | 2893 | 3223 | 3478 | 10618 | 9.62 | **51.3%** |
| | | OA | 63 | 5.4% | 89 | 220 | 273 | 335 | 917 | 14.56 | |
| International Journal of Epidemiology | 90 | Paywalled | 176 | 72.1% | 291 | 820 | 1270 | 1500 | 3881 | 22.05 | **91.7%** |
| | | OA | 68 | 27.9% | 187 | 575 | 893 | 1220 | 2875 | 42.28 | |
| International Journal of Oncology | 74 | Paywalled | 273 | 69.5% | 195 | 750 | 890 | 1094 | 2929 | 10.73 | **39.6%** |
| | | OA | 120 | 30.5% | 102 | 427 | 595 | 673 | 1797 | 14.98 | |
| Journal of Allergy and Clinical Immunology | 93 | Paywalled | 394 | 85.3% | 1241 | 2946 | 3189 | 3393 | 10769 | 27.33 | **17.2%** |
| | | OA | 68 | 14.7% | 276 | 582 | 654 | 667 | 2179 | 32.04 | |
| Journal of Antimicrobial Chemotherapy | 95 | Paywalled | 439 | 89.8% | 612 | 1878 | 2326 | 2393 | 7209 | 16.42 | **47.6%** |
| | | OA | 50 | 10.2% | 75 | 288 | 438 | 411 | 1212 | 24.24 | |
| Journal of Clinical Endocrinology and Metabolism | 99 | Paywalled | 407 | 88.9% | 438 | 1987 | 2582 | 2746 | 7753 | 19.05 | **-3.2%** |
| | | OA | 51 | 11.1% | 40 | 249 | 324 | 328 | 941 | 18.45 | |
| Journal of Clinical Microbiology | 86 | Paywalled | 269 | 81.0% | 291 | 1199 | 1491 | 1486 | 4467 | 16.61 | **8.0%** |
| | | OA | 63 | 19.0% | 94 | 342 | 371 | 323 | 1130 | 17.94 | |
| Journal of Epidemiology and Community Health | 94 | Paywalled | 95 | 66.9% | 58 | 239 | 355 | 402 | 1054 | 11.09 | **49.4%** |
| | | OA | 47 | 33.1% | 46 | 180 | 232 | 321 | 779 | 16.57 | |
| Journal of Infectious Diseases | 95 | Paywalled | 369 | 76.9% | 508 | 1648 | 2129 | 2246 | 6531 | 17.70 | **-20.1%** |
| | | OA | 111 | 23.1% | 162 | 427 | 527 | 520 | 1636 | 14.74 | |
| Journal of Investigative Dermatology | 98 | Paywalled | 280 | 87.2% | 418 | 1043 | 1319 | 1479 | 4259 | 15.21 | **11.6%** |
| | | OA | 41 | 12.8% | 54 | 196 | 199 | 247 | 696 | 16.98 | |
| Journal of Neurology | 84 | Paywalled | 204 | 82.6% | 188 | 591 | 767 | 921 | 2467 | 12.09 | **27.9%** |
| | | OA | 43 | 17.4% | 45 | 155 | 216 | 249 | 665 | 15.47 | |
| Journal of the American College of Cardiology | 99 | Paywalled | 235 | 85.1% | 1184 | 4061 | 4837 | 5286 | 15368 | 65.40 | **48.3%** |
| | | OA | 41 | 14.9% | 292 | 1045 | 1159 | 1479 | 3975 | 96.95 | |
| Lancet | 99 | Paywalled | 137 | 72.9% | 1904 | 6006 | 7701 | 8435 | 24046 | 175.52 | **96.0%** |
| | | OA | 51 | 27.1% | 829 | 3994 | 5907 | 6819 | 17549 | 344.10 | |
| Leukemia | 97 | Paywalled | 191 | 78.3% | 602 | 1305 | 1346 | 1473 | 4726 | 24.74 | **47.5%** |
| | | OA | 53 | 21.7% | 202 | 529 | 603 | 600 | 1934 | 36.49 | |
| Magnetic Resonance In Medicine | 92 | Paywalled | 419 | 88.4% | 597 | 1443 | 1677 | 1617 | 5334 | 12.73 | **17.1%** |
| | | OA | 55 | 11.6% | 104 | 228 | 243 | 245 | 820 | 14.91 | |
| Molecular Medicine Reports | 39 | Paywalled | 904 | 49.7% | 199 | 1197 | 1646 | 2150 | 5192 | 5.74 | **26.0%** |
| | | OA | 915 | 50.3% | 250 | 1508 | 2232 | 2634 | 6624 | 7.24 | |
| Molecular Psychiatry | 99 | Paywalled | 113 | 70.6% | 397 | 862 | 1041 | 1136 | 3436 | 30.41 | **63.9%** |
| | | OA | 47 | 29.4% | 283 | 592 | 714 | 754 | 2343 | 49.85 | |
| Oncology Reports | 65 | Paywalled | 609 | 74.4% | 431 | 1595 | 2104 | 2555 | 6685 | 10.98 | **21.0%** |
| | | OA | 210 | 25.6% | 108 | 656 | 917 | 1108 | 2789 | 13.28 | |
| Pharmaceutical Biology | 79 | Paywalled | 224 | 81.2% | 161 | 549 | 718 | 879 | 2307 | 10.30 | **-4.8%** |
| | | OA | 52 | 18.8% | 56 | 134 | 157 | 164 | 511 | 9.83 | |
| Psychological Medicine | 97 | Paywalled | 176 | 81.9% | 204 | 664 | 938 | 1166 | 2972 | 16.89 | **26.6%** |
| | | OA | 39 | 18.1% | 41 | 210 | 253 | 330 | 834 | 21.38 | |
| Quality of Life Research | 86 | Paywalled | 240 | 85.1% | 116 | 424 | 662 | 764 | 1966 | 8.19 | **6.4%** |
| | | OA | 42 | 14.9% | 16 | 64 | 118 | 168 | 366 | 8.71 | |
| Supportive Care in Cancer | 61 | Paywalled | 333 | 88.6% | 240 | 779 | 948 | 1189 | 3156 | 9.48 | **9.4%** |
| | | OA | 43 | 11.4% | 31 | 97 | 139 | 179 | 446 | 10.37 | |
| Surgical Endoscopy | 91 | Paywalled | 604 | 92.9% | 494 | 1680 | 2017 | 2638 | 6829 | 11.31 | **30.6%** |
| | | OA | 46 | 7.1% | 51 | 170 | 185 | 273 | 679 | 14.76 | |
| Vaccine | 99 | Paywalled | 631 | 73.4% | 366 | 1512 | 1853 | 2001 | 5732 | 9.08 | **38.8%** |
| | | OA | 229 | 26.6% | 198 | 811 | 937 | 942 | 2888 | 12.61 | |
| **Physics and Astronomy** | | | | | | | | | | | |
| 2d Materials | 95 | Paywalled | 244 | 82.4% | 552 | 1486 | 1764 | 1881 | 5683 | 23.29 | **39.0%** |
| | | OA | 52 | 17.6% | 155 | 463 | 548 | 517 | 1683 | 32.37 | |
| ACS Nano | 99 | Paywalled | 1228 | 95.1% | 2938 | 15752 | 20346 | 20317 | 59353 | 48.33 | **-42.5%** |
| | | OA | 63 | 4.9% | 108 | 514 | 738 | 777 | 2137 | 33.92 | |
| ACS Photonics | 95 | Paywalled | 371 | 95.4% | 525 | 2393 | 2864 | 2797 | 8579 | 23.12 | **3.1%** |
| | | OA | 18 | 4.6% | 14 | 119 | 153 | 143 | 429 | 23.83 | |
| Advanced Functional Materials | 97 | Paywalled | 779 | 96.9% | 2219 | 11199 | 14744 | 14737 | 42899 | 55.07 | **-46.3%** |
| | | OA | 25 | 3.1% | 56 | 241 | 304 | 340 | 941 | 37.64 | |
| Advanced Optical Materials | 95 | Paywalled | 288 | 94.7% | 402 | 1814 | 2404 | 2444 | 7064 | 24.53 | **-7.8%** |
| | | OA | 16 | 5.3% | 30 | 103 | 121 | 110 | 364 | 22.75 | |
| Applied Physics B Lasers and Optics | 78 | Paywalled | 248 | 86.4% | 131 | 444 | 519 | 435 | 1529 | 6.17 | **71.8%** |
| | | OA | 39 | 13.6% | 43 | 120 | 129 | 121 | 413 | 10.59 | |
| Applied Physics Express | 92 | Paywalled | 310 | 89.6% | 230 | 909 | 1146 | 1048 | 3333 | 10.75 | **-15.5%** |
| | | OA | 36 | 10.4% | 26 | 90 | 115 | 104 | 335 | 9.31 | |
| Astronomical Journal | 78 | Paywalled | 516 | 96.8% | 781 | 2803 | 2558 | 2392 | 8534 | 16.54 | **52.9%** |
| | | OA | 17 | 3.2% | 23 | 103 | 143 | 161 | 430 | 25.29 | |
| Astrophysical Journal Letters | 93 | Paywalled | 513 | 94.6% | 1167 | 4133 | 4504 | 3953 | 13757 | 26.82 | **307.5%** |
| | | OA | 29 | 5.4% | 163 | 972 | 1137 | 897 | 3169 | 109.28 | |
| Chaos | 86 | Paywalled | 349 | 94.3% | 293 | 1007 | 1258 | 1138 | 3696 | 10.59 | **-18.3%** |
| | | OA | 21 | 5.7% | 18 | 47 | 63 | 60 | 188 | 8.95 | |
| Chinese Physics C | 98 | Paywalled | 163 | 74.1% | 145 | 455 | 513 | 583 | 1696 | 10.40 | **-31.2%** |
| | | OA | 57 | 25.9% | 64 | 147 | 124 | 117 | 452 | 7.93 | |

| Journal | | Type | | | | | | | | | |
|---|---|---|---|---|---|---|---|---|---|---|---|
| Communications in Mathematical Physics | 99 | Paywalled | 235 | 89.4% | 208 | 504 | 593 | 696 | 2001 | 8.51 | 27.5% |
| | | OA | 28 | 10.6% | 28 | 84 | 88 | 104 | 304 | 10.86 | |
| European Physical Journal B | 52 | Paywalled | 235 | 91.8% | 64 | 290 | 312 | 310 | 976 | 4.15 | 44.5% |
| | | OA | 21 | 8.2% | 14 | 35 | 43 | 34 | 126 | 6.00 | |
| European Physical Journal D | 47 | Paywalled | 301 | 91.8% | 153 | 400 | 389 | 370 | 1312 | 4.36 | -9.0% |
| | | OA | 27 | 8.2% | 18 | 37 | 31 | 22 | 108 | 4.00 | |
| Experiments in Fluids | 92 | Paywalled | 136 | 82.9% | 50 | 283 | 431 | 419 | 1183 | 8.70 | 33.8% |
| | | OA | 28 | 17.1% | 23 | 90 | 95 | 118 | 326 | 11.64 | |
| Japanese Journal of Applied Physics | 76 | Paywalled | 681 | 96.3% | 212 | 1035 | 1009 | 971 | 3227 | 4.74 | 43.7% |
| | | OA | 26 | 3.7% | 13 | 47 | 60 | 57 | 177 | 6.81 | |
| Journal of Chemical Physics | 88 | Paywalled | 1952 | 97.0% | 1490 | 5331 | 5794 | 5381 | 17996 | 9.22 | 35.4% |
| | | OA | 60 | 3.0% | 55 | 211 | 241 | 242 | 749 | 12.48 | |
| Journal of Computational Physics | 85 | Paywalled | 684 | 95.3% | 430 | 1807 | 2748 | 2816 | 7801 | 11.40 | 16.0% |
| | | OA | 34 | 4.7% | 29 | 110 | 152 | 159 | 450 | 13.24 | |
| Journal of Fluid Mechanics | 89 | Paywalled | 736 | 93.3% | 568 | 2083 | 3053 | 3211 | 8915 | 12.11 | 16.5% |
| | | OA | 53 | 6.7% | 53 | 201 | 261 | 233 | 748 | 14.11 | |
| Journal of Instrumentation | 68 | Paywalled | 353 | 82.5% | 124 | 483 | 542 | 492 | 1641 | 4.65 | 327.6% |
| | | OA | 75 | 17.5% | 148 | 494 | 489 | 360 | 1491 | 19.88 | |
| Journal of Magnetic Resonance | 83 | Paywalled | 183 | 90.6% | 135 | 462 | 508 | 411 | 1516 | 8.28 | 72.2% |
| | | OA | 19 | 9.4% | 13 | 66 | 98 | 94 | 271 | 14.26 | |
| Journal of Nanoparticle Research | 80 | Paywalled | 372 | 94.9% | 119 | 593 | 812 | 733 | 2257 | 6.07 | 157.1% |
| | | OA | 20 | 5.1% | 9 | 62 | 116 | 125 | 312 | 15.60 | |
| Journal of Nuclear Materials | 87 | Paywalled | 531 | 96.4% | 266 | 1302 | 1552 | 1604 | 4724 | 8.90 | 19.7% |
| | | OA | 20 | 3.6% | 13 | 49 | 76 | 75 | 213 | 10.65 | |
| Journal of Optics United Kingdom | 70 | Paywalled | 336 | 91.8% | 204 | 717 | 807 | 720 | 2448 | 7.29 | -1.7% |
| | | OA | 30 | 8.2% | 23 | 69 | 61 | 62 | 215 | 7.17 | |
| Journal of Physics A Mathematical and Theoretical | 91 | Paywalled | 630 | 97.4% | 546 | 1199 | 1203 | 1153 | 4101 | 6.51 | 28.3% |
| | | OA | 17 | 2.6% | 12 | 51 | 36 | 43 | 142 | 8.35 | |
| Journal of Physics Condensed Matter | 80 | Paywalled | 770 | 96.4% | 454 | 1589 | 1925 | 2198 | 6166 | 8.01 | 16.3% |
| | | OA | 29 | 3.6% | 28 | 81 | 90 | 71 | 270 | 9.31 | |
| Journal of Physics D Applied Physics | 83 | Paywalled | 1095 | 94.4% | 685 | 2525 | 3043 | 2890 | 9143 | 8.35 | 51.3% |
| | | OA | 65 | 5.6% | 80 | 229 | 250 | 262 | 821 | 12.63 | |
| Journal of Physics G Nuclear and Particle Physics | 87 | Paywalled | 129 | 80.6% | 115 | 336 | 281 | 317 | 1049 | 8.13 | 15.4% |
| | | OA | 31 | 19.4% | 16 | 55 | 99 | 121 | 291 | 9.39 | |
| Journal of Quant Spectroscopy and Radiative Transfer | 84 | Paywalled | 391 | 95.6% | 503 | 1097 | 1090 | 1149 | 3839 | 9.82 | 685.9% |
| | | OA | 18 | 4.4% | 54 | 309 | 472 | 554 | 1389 | 77.17 | |
| Journal of Statistical Physics | 73 | Paywalled | 229 | 92.3% | 94 | 323 | 366 | 379 | 1162 | 5.07 | 16.2% |
| | | OA | 19 | 7.7% | 18 | 32 | 25 | 37 | 112 | 5.89 | |
| Journal of Synchrotron Radiation | 95 | Paywalled | 76 | 69.1% | 36 | 124 | 121 | 156 | 437 | 5.75 | 74.4% |
| | | OA | 34 | 30.9% | 36 | 87 | 106 | 112 | 341 | 10.03 | |
| Journal of the Acoustical Society of America | 71 | Paywalled | 805 | 93.2% | 322 | 1364 | 1850 | 1642 | 5178 | 6.43 | 24.6% |
| | | OA | 59 | 6.8% | 28 | 120 | 167 | 158 | 473 | 8.02 | |
| Journal of the Physical Society of Japan | 63 | Paywalled | 369 | 87.0% | 152 | 481 | 508 | 390 | 1531 | 4.15 | 89.3% |
| | | OA | 55 | 13.0% | 27 | 149 | 144 | 112 | 432 | 7.85 | |
| Journal of Thermal Analysis and Calorimetry | 70 | Paywalled | 793 | 93.3% | 467 | 1841 | 1954 | 2099 | 6361 | 8.02 | 9.8% |
| | | OA | 57 | 6.7% | 40 | 145 | 161 | 156 | 502 | 8.81 | |
| Measurement Science and Technology | 77 | Paywalled | 460 | 95.2% | 227 | 803 | 967 | 937 | 2934 | 6.38 | 29.5% |
| | | OA | 23 | 4.8% | 17 | 56 | 64 | 53 | 190 | 8.26 | |
| Nano Letters | 99 | Paywalled | 1086 | 96.1% | 2637 | 12267 | 14875 | 14776 | 44555 | 41.03 | -61.2% |
| | | OA | 44 | 3.9% | 64 | 337 | 367 | 352 | 1120 | 25.45 | |
| Nuclear Fusion | 87 | Paywalled | 467 | 91.4% | 637 | 1663 | 1723 | 1518 | 5541 | 11.87 | 84.8% |
| | | OA | 44 | 8.6% | 87 | 270 | 295 | 313 | 965 | 21.93 | |
| Nuclear Physics A | 75 | Paywalled | 144 | 41.9% | 192 | 332 | 337 | 329 | 1190 | 8.26 | -119.2% |
| | | OA | 200 | 58.1% | 47 | 281 | 242 | 184 | 754 | 3.77 | |
| Optical and Quantum Electronics | 52 | Paywalled | 397 | 94.1% | 217 | 821 | 787 | 799 | 2624 | 6.61 | -53.0% |
| | | OA | 25 | 5.9% | 18 | 33 | 37 | 20 | 108 | 4.32 | |
| Optical Engineering | 74 | Paywalled | 616 | 88.3% | 165 | 650 | 830 | 721 | 2366 | 3.84 | 105.4% |
| | | OA | 82 | 11.7% | 61 | 165 | 209 | 212 | 647 | 7.89 | |
| Physical Review Applied | 91 | Paywalled | 402 | 95.7% | 403 | 1733 | 2041 | 2058 | 6235 | 15.51 | -71.3% |
| | | OA | 18 | 4.3% | 9 | 50 | 60 | 44 | 163 | 9.06 | |
| Physical Review C | 89 | Paywalled | 1008 | 96.9% | 983 | 3151 | 3145 | 3109 | 10388 | 10.31 | 33.7% |
| | | OA | 32 | 3.1% | 45 | 145 | 133 | 118 | 441 | 13.78 | |
| Physics of Fluids | 74 | Paywalled | 600 | 97.2% | 260 | 1500 | 2365 | 2117 | 6242 | 10.40 | 18.2% |
| | | OA | 17 | 2.8% | 9 | 46 | 88 | 66 | 209 | 12.29 | |
| Physics of Plasmas | 69 | Paywalled | 1317 | 96.3% | 720 | 2414 | 2491 | 2375 | 8000 | 6.07 | 97.2% |
| | | OA | 51 | 3.7% | 35 | 176 | 203 | 197 | 611 | 11.98 | |
| Quantum Information Processing | 79 | Paywalled | 294 | 93.9% | 154 | 593 | 921 | 813 | 2481 | 8.44 | -51.3% |
| | | OA | 19 | 6.1% | 10 | 22 | 34 | 40 | 106 | 5.58 | |
| Review of Scientific Instruments | 68 | Paywalled | 832 | 94.2% | 271 | 1256 | 1499 | 1509 | 4535 | 5.45 | 54.3% |
| | | OA | 51 | 5.8% | 35 | 101 | 143 | 150 | 429 | 8.41 | |
| Soft Matter | 93 | Paywalled | 2335 | 95.6% | 591 | 2256 | 2645 | 2638 | 8130 | 3.48 | 247.3% |
| | | OA | 107 | 4.4% | 105 | 379 | 408 | 402 | 1294 | 12.09 | |
| Solar Physics | 74 | Paywalled | 166 | 87.4% | 131 | 370 | 406 | 412 | 1319 | 7.95 | 41.1% |
| | | OA | 24 | 12.6% | 30 | 90 | 78 | 71 | 269 | 11.21 | |
| Superconductor Science and Technology | 90 | Paywalled | 259 | 93.5% | 215 | 590 | 806 | 738 | 2349 | 9.07 | 18.8% |
| | | OA | 18 | 6.5% | 13 | 48 | 56 | 77 | 194 | 10.78 | |
| Ultramicroscopy | 89 | Paywalled | 209 | 92.5% | 156 | 453 | 492 | 451 | 1552 | 7.43 | 172.5% |
| | | OA | 17 | 7.5% | 33 | 84 | 112 | 115 | 344 | 20.24 | |